\documentclass[a4paper,11pt]{article}
\usepackage{jcappub,tablefootnote,adjustbox}  
\usepackage{lineno}
\usepackage{cancel,ulem}\usepackage{xcolor}

\newcommand{\bea}{\begin{eqnarray}}
 
\newcommand{\eea}{\end{eqnarray}}
\newcommand{\bma}{\begin{pmatrix}}
\newcommand{\ema}{\end{pmatrix}}
\newcommand{\be}{\begin{equation}}
\newcommand{\ee}{\end{equation}}
\newcommand{\beno}{\begin{equation*}}
\newcommand{\eeno}{\end{equation*}}

\def\doi{http://doi.org}

\newcommand{\m}{\mathring}

\title{Non-metricity with boundary terms:  $f(Q,C)$ gravity and cosmology }

\author[a]{Avik De,}
\affiliation[a]{Department of Mathematical and Actuarial Sciences\\ Universiti Tunku Abdul Rahman, Jalan Sungai Long,
43000 Cheras, Malaysia}
\author[b]{Tee-How Loo}
\affiliation[b]{Institute of Mathematical Sciences, Faculty of Science, Universiti Malaya, 50603 Kuala Lumpur, Malaysia}
\author[c]{and Emmanuel N. Saridakis}
\affiliation[c]{National Observatory of Athens, Lofos Nymfon, 11852 Athens, 
Greece}
\affiliation[c]{CAS Key Laboratory for Researches in Galaxies and Cosmology, 
Department of Astronomy, University of Science and Technology of China, Hefei, 
Anhui 230026, P.R. China}
 \affiliation[c]{Departamento de Matem\'{a}ticas, Universidad Cat\'{o}lica del 
Norte, Avda.
Angamos 0610, Casilla 1280 Antofagasta, Chile}

\emailAdd{avikde@utar.edu.my, looth@um.edu.my, msaridak@noa.gr}

\abstract{We formulate $f(Q,C)$ gravity and cosmology. Such a 
construction   is based on the symmetric teleparallel geometry, but apart 
form the non-metricity scalar $Q$ we incorporate in the Lagrangian the 
boundary term $C$ of its difference form the standard Levi-Civita Ricci scalar 
$\m R$.  We extract the general metric and affine connection field equations,
 we apply them at a cosmological framework, and 
  adopting three different types of symmetric teleparallel affine connections 
  we   obtain the modified Friedmann equations. As we show, we acquire an 
effective dark-energy sector of geometrical origin, which can lead to 
interesting cosmological phenomenology.
Additionally, 
we may obtain an effective interaction between matter and dark energy.
  Finally, examining a specific model, we 
 show that  we can   obtain the usual thermal 
history of the universe, with the sequence of matter and dark-energy epochs, 
while  the   effective dark-energy equation-of-state 
parameter  can be quintessence-like, phantom-like, or cross the phantom-divide 
during evolution.
}

\begin{document}
\maketitle
\flushbottom



\section{Introduction}

In order to solve possible theoretical issues of general relativity, such as 
non-renormalizability and the cosmological constant problem, and    alleviate 
possible tensions between its predictions and the data 
\cite{Perivolaropoulos:2021jda,Abdalla:2022yfr}, many theories of gravitational 
modification have been proposed   \cite{CANTATA:2021ktz,Capozziello:2011et,lavinia:2019}.
 There are many directions one could follow to   construct gravitational 
modifications. For instance,  one can start from the standard curvature 
formulation of gravity and  result to    $f(R)$ gravity 
\cite{Starobinsky:1980te,DeFelice:2010aj},  to  $f(G)$
gravity \cite{Nojiri:2005jg}, to  $f(P)$ 
gravity 
\cite{Erices:2019mkd}, to Lovelock gravity \cite{Lovelock:1971yv}, etc. 
Alternatively, one can start from 
the torsional formulation of gravity and   obtain 
   $f(\mathbb{T})$ gravity 
\cite{Bengochea:2008gz,Cai:2015emx},   $f(\mathbb{T},\mathbb{T}_{G})$ gravity 
\cite{Kofinas:2014owa,Kofinas:2014daa},
etc.
 Finally, there is an alternative way to build classes of modified gravities, 
namely to use non-metricity   \cite{Nester,coincident}
resulting to $f(Q)$ gravity 
\cite{coincident,fQfT,Iosifidis:2018zjj,fQfT1,fQfT2,fQfT3,Iosifidis:2020zzp, 
Iosifidis:2020gth,dynamical1,
cosmo-fQ,lcdm,zhao,accfQ1,accfQ2,accfQ3,deepjc,gde,lin,cosmography,signa, 
redshift,perturb,dynamical2,latetime,quantum,bouncing,bigbang,Avik/prd,de/phase, 
de/probe,fQec1,fQec2,FLRW/connection1,FLRW/connection,bh, 
avik/cqg,23,40,Iosifidis:2023pvz,Shabani:2023xfn,Shi:2023kvu}.

The above novel classes of modified gravity in curvature, torsional and 
non-metricity case, arise although the non-modified theories are equivalent at 
the level of equations. The reason behind this is that 
 the  torsion scalar 
$\mathbb{T}$   and the non-metricity scalar $Q$  differ from the usual 
Levi-Civita Ricci scalar $\m R$ of general relativity by a 
total divergence term, namely  $
\m R=-\mathbb T+B$ and $
\m R=Q+C$ respectively, and thus   arbitrary functions $f(\m R)$, 
$f(\mathbb{T})$ and $f(Q)$ do not differ by a total derivative anymore.
Finally, note that one can also introduce scalar fields in the above framework, 
obtaining scalar - tensor  \cite{Horndeski:1974wa,DeFelice:2010nf}, scalar 
- torsion  \cite{Geng:2011aj,hor2,Bahamonde:2022cmz} and scalar - non-metricity 
\cite{41,42} 
theories.
 
In the framework of teleparallel gravities one may incorporate the boundary $B$ 
into the Lagrangian, resulting to   $f(\mathbb{T},B)$ theories
  \cite{Bahamonde:2015zma}, which as expected exhibit richer phenomenology   
\cite{teleparallel}. Nevertheless, in the framework of non-metricity the role 
of $C$ has not been   incorporated into the Lagrangian of   symmetric 
teleparallel gravity. \footnote{While this manuscript was being proofread, the 
work \cite{Capozziello:2023vne} appeared on arxiv with discussions on the role 
of the boundary 
term on non-metricity gravity,  namely $f(Q,B)$ gravity in their notation, but 
without cosmological applications. We agree on \cite{Capozziello:2023vne} in 
regions of 
overlap.} Hence, in this work we are interested in investigating 
such a direction, namely to  formulate   $f(Q,C)$ gravity and apply it to a 
cosmological framework.
 
The present article is organized as follows:  In Section 
\ref{sec0} we present the geometrical framework of symmetric teleparallelism. 
Then, in section \ref{sec1} we formulate $f(Q,C)$ gravity, extracting the 
general metric and affine connection field equations, while in section 
\ref{sec4} we apply it to a 
cosmological setup, thus obtaining $f(Q,C)$ cosmology.
 Finally  in section \ref{sec5} we conclude.


\section{Symmetric teleparallel geometries}\label{sec0}

Let us begin with a brief introduction on the general framework,
namely the teleparallel geometries.
In general, a metric-affine geometry consists of a $4$-dimensional Lorentzian 
manifold $M$, a line element governed by the metric tensor $g_{\mu\nu}$ in a
 coordinate system $\{x^0,x^1,x^2,x^3\}$ and a non-tensorial term, the 
affine connection $\Gamma^{\alpha}_{\,\,\,\mu\nu}$, defining the covariant 
derivative $\nabla_\lambda$. Although the metric and the connection are 
completely independent objects, if one   imposes both the 
metric-compatibility and torsion-free conditions, then
  there is   a unique connection available, namely the Levi-Civita 
connection $\mathring{\Gamma}^\alpha{}_{\mu\nu}$, in which case    it has a 
well-known relation with the metric $g$ given by
\begin{equation}
\mathring{\Gamma}^\alpha_{\,\,\,\mu\nu}=\frac{1}{2} 
g^{\alpha\beta}\left(\partial_\nu g_{\beta\mu}+\partial_\mu 
g_{\beta\nu}-\partial_\beta g_{\mu\nu}  \right).
\end{equation}
In such a cimple case  the triplet 
$(M,g_{\mu\nu},\Gamma^\alpha{}_{\mu\nu})$ constitutes the Riemannian geometry, 
which is the basis of general relativity.
 
 Nevertheless, things change if we start relaxing the aforementioned 
conditions. One direction is to maintain metric compatibility but introduce 
connections that have zero curvature but non-zero torsion, such as the  
Weitzenb\"{o}ck connection used in the Teleparallel Equivalent of General 
Relativity (TEGR) \cite{Cai:2015emx}. One other direction, 
is to consider 
a torsion-free and curvature-free affine connection $\Gamma^\alpha{}_{\mu\nu}$, 
namely with
\begin{align}
\mathbb T^\alpha{}_{\mu\nu}:=&
2\Gamma^{\alpha}{}_{[\nu\mu]}=0\,,
\label{eqn:torsion-free}\\
R^\lambda{}_{\mu\alpha\nu}:=&
2\partial_{[\alpha}\Gamma^\lambda{}_{|\mu|\nu]}
+2\Gamma^\lambda{}_{\sigma[\alpha}\Gamma^\sigma{}_{|\mu|\nu]}=0\,, 
\label{eqn:curvature-free}
\end{align}
however relaxing metric compatibility, resulting to a class of geometries called
 symmetric teleparallel geometries. 
 In particular, due to the disappearance of the Riemannian 
curvature tensor, the parallel transport defined by the covariant derivative and 
its associated affine connection is independent of the path, hence the term 
``teleparallel''. In addition, due to the torsionless constraint on the 
connection,   the affine connection is symmetric in its lower indices, 
hence the term ``symmetric''. 

The incompatibility of the above affine connection 
with the metric is quantified by the non-metricity tensor \cite{coincident} 
\begin{equation} \label{Q tensor}
Q_{\lambda\mu\nu} := \nabla_\lambda g_{\mu\nu}= \partial_\lambda 
g_{\mu\nu}-\Gamma^{\beta}_{\,\,\,\lambda\mu}g_{\beta\nu}-\Gamma^{\beta}_{\,\,\,
\lambda\nu}g_{\beta\mu}\neq 0 \,.
\end{equation}
Moreover, we can express
\begin{equation} \label{connc}
\Gamma^\lambda{}_{\mu\nu} := 
\mathring{\Gamma}^\lambda{}_{\mu\nu}+L^\lambda{}_{\mu\nu},
\end{equation}
where $L^\lambda{}_{\mu\nu}$ is the disformation tensor. Thus, it follows that
\begin{equation} \label{L}
L^\lambda{}_{\mu\nu} = \frac{1}{2} (Q^\lambda{}_{\mu\nu} - Q_\mu{}^\lambda{}_\nu - Q_\nu{}^\lambda{}_\mu) \,.
\end{equation}
We can now construct two different types of non-metricity vectors, i.e.
\begin{eqnarray}
 Q_\mu := g^{\nu\lambda}Q_{\mu\nu\lambda} = Q_\mu{}^\nu{}_\nu \,, \\
 \tilde{Q}_\mu 
:= g^{\nu\lambda}Q_{\nu\mu\lambda} = Q_{\nu\mu}{}^\nu \,,
\end{eqnarray}
and similarly  we can write
\begin{eqnarray}
 L_\mu := L_\mu{}^\nu{}_\nu \,, \\
 \tilde{L}_\mu := L_{\nu\mu}{}^\nu \,.   
\end{eqnarray}
Furthermore, the superpotential (or the non-metricity conjugate) tensor 
$P^\lambda{}_{\mu\nu}$ is given by
\begin{equation} \label{P}
P^\lambda{}_{\mu\nu} = 
\frac{1}{4} \left[ -2 L^\lambda{}_{\mu\nu} + Q^\lambda g_{\mu\nu} - 
\tilde{Q}^\lambda g_{\mu\nu} -\delta^\lambda{}_{(\mu} Q_{\nu)} \right] \,,
\end{equation}
while the non-metricity scalar $Q$ is defined as
\begin{equation} \label{Q}
Q=Q_{\alpha\beta\gamma}P^{\alpha\beta\gamma}\,.
\end{equation}

In summary, after introducing the torsion-free and curvature-free constraints 
(\ref{eqn:torsion-free}),(\ref{eqn:curvature-free}),  one can further   
obtain the   relations (all quantities with a $\mathring{(~)}$ are 
calculated with respect to the 
Levi-Civita connection $\mathring{\Gamma}$):
\begin{equation}
 \m R_{\mu\nu}+\m\nabla_\alpha L^\alpha{}_{\mu\nu}-\m\nabla_\nu\tilde L_\mu
+\tilde L_\alpha L^\alpha{}_{\mu\nu}-L_{\alpha\beta\nu}L^{\beta\alpha}{}_\mu=0,
\end{equation}
\begin{equation}
\label{mRicci}
 \m R+\m\nabla_\alpha(L^\alpha-\tilde L^\alpha)-Q=0\,.
\end{equation}
Hence, it becomes clear that since $Q^\alpha-\tilde Q^\alpha=L^\alpha-\tilde 
L^\alpha$, 
from the preceding relation we also define the boundary term   
\begin{eqnarray}
&&
\!\!\!\!\!\!\!\!
C:=\m{R}-Q=-\m\nabla_\alpha(Q^\alpha-\tilde Q^\alpha)\notag\\
&&=-\frac1{\sqrt{-g}}\partial_\alpha\left[\sqrt{-g}(Q^\alpha-\tilde 
Q^\alpha)\right].
\label{Cbounday}
\end{eqnarray}


\section{$f(Q,C)$ gravity}\label{sec1}

In the previous section we presented the symmetric teleparallel geometry. In 
this section we will show how one can use it as the framework to construct a 
new theory of gravity, namely $f(Q,C)$ gravity.

General relativity is constructed in the framework of Riemannian geometry, 
and the gravitational action, 
the Einstein-Hilbert action, is  
\begin{equation}
S_{GR}=\int \frac{1}{2\kappa }\m R \sqrt{-g}%
\,d^{4}x\,,
\end{equation}
with $\m R$ the Ricci scalar calculated with the Levi-Civita connection.
Similarly, Teleparallel Equivalent of general relativity (TEGR)
is constructed in the framework of Weitzenb\"{o}ck  geometry, 
and the gravitational action 
  is  
\begin{equation}
S_{TEGR}=\int \frac{1}{2\kappa } \mathbb T \m R
\,d^{4}x\,,
\end{equation}
with $\mathbb T$ the torsion scalar calculated with the  Weitzenb\"{o}ck  
connection, and $e = \text{det}(e_{\mu}^A) = \sqrt{-g}$.
Hence, in the same lines,   Symmetric Teleparallel   general relativity (STGR)
is constructed in the framework of symmetric teleparallel geometry, 
and the gravitational action  is given by
\begin{equation}
S_{STGR}=\int \frac{1}{2\kappa }Q \sqrt{-g}\,
 d^{4}x\,,
\end{equation}
where the non-metricity scalar $Q$ is  calculated with the  symmetric 
teleparallel connection and it is given in (\ref{Q}). All these theories are 
completely equivalent at the level of equations, since their Lagrangians 
differ only by boundary terms.

As it is known, one can start from  the above three gravitational formulations, 
and construct modifications, resulting to  $f(\m R)$ gravity 
\cite{DeFelice:2010aj}, to  $f(\mathbb T)$ gravity \cite{Cai:2015emx}, or to 
$f(Q)$ gravity    with action 
\cite{coincident, dynamical1, cosmo-fQ, lcdm, zhao}
\begin{equation}
S=\int \frac{1}{2\kappa }f(Q) \sqrt{-g}%
\,d^{4}x\,.
\end{equation}
However, as expected, the resulting modified theories of gravity, namely $f(\m 
R)$, $f(\mathbb T)$   and $f(Q)$ ones, are not equivalent  since their 
differences are not boundary terms any more. Finally,
Since $\m 
R=- \mathbb T+B$, where 
\begin{equation}
 B=2\m \nabla_\mu {\mathbb T}_{\sigma}^{\,\,\,\,\sigma\mu},
\end{equation}
is the boundary term, one could 
construct an even richer gravitational modification, namely  $f(\mathbb T,B)$ 
gravity  \cite{Bahamonde:2015zma} with interesting cosmological phenomenology 
\cite{Bahamonde:2015zma,Farrugia:2018gyz,Escamilla-Rivera:2019ulu,
Paliathanasis:2017efk,Paliathanasis:2017flf,Paliathanasis:2021kuh,
Paliathanasis:2021ysb,Paliathanasis:2022pgu,
Caruana:2020szx, Moreira:2021xfe}, which in 
the case $f(\mathbb T,B)=f(-\mathbb T+B)$ coincides with $f(\m 
R)$ gravity.

In this work we proceed to the construction of a new theory of gravity based on 
symmetric teleparallel geometry, but incorporating both $Q$ and the boundary 
$C$ of (\ref{Cbounday}) in the Lagrangian, namely  $f(Q,C)$ gravity. Hence, we 
write the action as 
\begin{equation}
S=\int \left[ \frac{1}{2\kappa }f(Q,C)\right] \sqrt{-g}%
\,d^{4}x\,,
\label{eqn:action-fQC}
\end{equation}
where $f$ is an arbitrary function on both $Q$ and $C$. Variation of the action 
with respect to the metric  (see   Appendix \ref{derive} for the details), 
adding also the matter Lagrangian a $\mathcal L_m$  for completeness, gives rise 
to the field equations
\begin{eqnarray}
&&
\!\!\!\!\!\!\!\!\!\!\!\!\!\!\!\!
\kappa T_{\mu\nu}
=-\frac f2g_{\mu\nu}
  +\frac2{\sqrt{-g}}\partial_\lambda \left(\sqrt{-g}f_Q P^\lambda{}_{\mu\nu}
  \right) 
  +(P_{\mu\alpha\beta}Q_\nu{}^{\alpha\beta}-2P_{\alpha\beta\nu}Q^{\alpha\beta}{}_\mu)
        f_Q \nonumber\\
&& \ \ 
  +\left(\frac C2 g_{\mu\nu}-\m\nabla_{\mu}\m\nabla_{\nu} 
+g_{\mu\nu}\m\nabla^\alpha\m\nabla_\alpha-2P^\lambda{}_{\mu\nu}\partial_\lambda 
\right)\!f_C,
\label{eqn:FE1-pre}
\end{eqnarray}
where $T_{\mu\nu}$ is the matter energy momentum tensor, and where  
$f_Q:=\frac{\partial f}{\partial Q}$ and $ f_{C}:=\frac{\partial f}{\partial 
C}$. We mention here that  since the affine connection is indepedent of the 
metric tensor, by performing variation of the action with respect to the affine 
connection  
we obtain the connection field equation (see
Appendix  \ref{derive}): 
\begin{align}\label{eqn:FE2-invar}
(\nabla_\mu-\tilde L_\mu)(\nabla_\nu-\tilde L_\nu)
\left[4(f_Q-f_C)P^{\mu\nu}{}_\lambda+\Delta_\lambda{}^{\mu\nu}\right]=0\,,
\end{align}
where 
$\Delta_\lambda{}^{\mu\nu}=-\frac2{\sqrt{-g}}
\frac{\delta(\sqrt{-g}
\mathcal L_m)}{\delta\Gamma^\lambda{}_{\mu\nu}}$
is the 
hypermomentum tensor
\cite{hyper}.
As 
\begin{align}\label{del:g}
\partial_\nu\sqrt{-g}=-\sqrt{-g}\tilde L_\nu\,,
\end{align}
while taking the coincident gauge, the preceding connection field equation can be re-expressed as 
\begin{align}\label{eqn:FE2b}
\partial_\mu\partial_\nu\left(\sqrt{-g}
\left[4(f_Q-f_C)P^{\mu\nu}{}_\lambda+\Delta_\lambda{}^{\mu\nu}\right]
\right)=0\,,
\end{align}
in which case the field equation  similar to that of $f(Q)$ gravity 
will be recovered \cite{harko-coupling}. See also the analysis of  
 \cite{Boehmer:2021aji,Boehmer:2023fyl} for useful relations.

Let us make some comments here. In the literature of symmetric teleparallel 
theories, the so-called ``coincident gauge'' is frequently used to designate a 
coordinate system in which the connection disappears and covariant derivatives 
reduce to partial derivatives (see the discussion in 
\cite{BeltranJimenez:2022azb}). As described in 
\cite{zhao}, this occasionally 
poses a significant problem when attempting to investigate other spacetimes 
using the same vanishing affine connections. In most cases, the
symmetries of the system make it inconsistent unless the non-metricity scalar 
$Q$ is forced to be a constant. Hence, one concludes that a fully-covariant 
formulation would be very useful for incorporating non-vanishing connections. 

As we observe from the  field equation  (\ref{eqn:FE1-pre}), the second and 
third terms on the right-hand side  
 constitute the $f(Q)$ theory. 
Following the standard calculation (for instance, see \cite{fQT,zhao}), we 
obtain
\begin{equation}
 \frac2{\sqrt{-g}}\partial_\lambda \left(\sqrt{-g}f_Q P^\lambda{}_{\mu\nu}
  \right)  
+(P_{\mu\alpha\beta}Q_\nu{}^{\alpha\beta}-2P_{\alpha\beta\nu}Q^{\alpha\beta}{}
_\mu)f_Q
  =\left(2P^\lambda{}_{\mu\nu}\partial_\lambda+\m G_{\mu\nu}+\frac Q2g_{\mu\nu}\right)f_Q\,,
 \end{equation}
where $\m G_{\mu\nu}$ is the Einstein tensor corresponding to the Levi-Civita 
connection.
Hence, we can rewrite the metric field equation covariantly as 
\begin{equation}\label{eqn:FE1}
\!\!\!\!\!
\kappa T_{\mu\nu}
=-\frac 
f2g_{\mu\nu}\!+\!2P^\lambda{}_{\mu\nu}\nabla_\lambda(f_Q\!-\!f_C)+\!\left(\!\m 
G_{\mu\nu}\!+\!\frac Q2g_{\mu\nu}\!\right)f_Q
  +\left(\!\frac C2g_{\mu\nu}\!-\!\m\nabla_{\mu}\m\nabla_{\nu}
  \!+\!g_{\mu\nu}\m\nabla^\alpha\m\nabla_\alpha \!\right)f_C\,.
 \end{equation}
As a next step, we define the effective stress energy tensor as
\begin{eqnarray} \label{T^eff}
 &&
 \!\!\!\!\! \!\!\!\!\! \!\!\!\!\! \!\!
 T^{\text{eff}}_{\mu\nu} =  T_{\mu\nu}+ \frac 1{\kappa}\left[\frac 
f2g_{\mu\nu}-2P^\lambda{}_{\mu\nu}\nabla_\lambda(f_Q-f_C)
 \right.
   \nonumber\\
&&
  \left.\!
-\frac {Qf_Q}2g_{\mu\nu}
-\left(\frac C2g_{\mu\nu}-\m\nabla_{\mu}\m\nabla_{\nu}
  +g_{\mu\nu}\m\nabla^\alpha\m\nabla_\alpha \right)f_C\right],
\end{eqnarray}
and thus we result to
\begin{align}
    \m G_{\mu\nu}=\frac{\kappa}{f_Q}T^{\text{eff}}_{\mu\nu}\,.
\end{align}
Hence, in the framework of $f(Q,C)$ gravity we obtain an extra, effective 
sector of geometrical origin.
 
If the  function $f$   is linear in $C$,  namely $f(Q,C)=f(Q)+\beta C$ and thus
$f_C=const.$, then     (\ref{eqn:FE1}) reduces to the 
usual field equation for $(Q)$ gravity:
\begin{align}
\kappa T_{\mu\nu}
&=-\frac f2g_{\mu\nu}+2P^\lambda{}_{\mu\nu}\nabla_\lambda f_Q
  +\left(\m G_{\mu\nu}+\frac Q2g_{\mu\nu}\right)f_Q\,.
\end{align}
This is the unique form of the function  $f$ which yields second-order field 
equations. On the other hand, the generic field equations (\ref{eqn:FE1}) 
contain terms of the form $\partial_\mu\partial_\nu f_C$ which are  of  
fourth order. Additionally, since $\m R=Q+C$, in order to recover $f(\m R)$ 
theory  we consider  
$ 
f(Q,C)=f(Q+C)$, in which case we acquire
\begin{align}
\kappa T_{\mu\nu}
&=-\frac f2g_{\mu\nu}
  +\left(\m R_{\mu\nu}-\m\nabla_{\mu}\m\nabla_{\nu}
  +g_{\mu\nu}\m\nabla^\alpha\m\nabla_\alpha \right)f_{\m R}\,,
\end{align}
which are the usual field equation for $f(\m R)$ gravity theory.

 We close this section with a discussion on the energy conservation in $f(Q,C)$ 
theories of gravity. In general relativity, as well as in $f(\m R)$ theories of 
gravity, the field equations are   compatible with the classical 
energy conservation   due to the second Bianchi identity, which is not 
the case in simple   $f(Q)$ gravity \cite{avik/cqg}.  On the other hand, in the 
   covariant formulation of $f(Q,C)$ gravity, equation (\ref{eqn:FE1}) leads to
\begin{eqnarray}\label{vv0}
&& 
\!\!\!\!\!\!\!\!\!\!\!\!\!\!\!\!\!\!\!
\kappa\m\nabla_\mu T^\mu{}_\nu
=-\frac{\nabla_\nu f}2+2\m\nabla_\mu P^{\lambda\mu}{}_\nu\nabla_\lambda(f_Q-f_C)
 +2P^{\lambda\mu}{}_\nu\m\nabla_\mu\nabla_\lambda(f_Q-f_C) \notag\\
&&
\ \ \ \ \ 
+\left(\m G^\lambda{}_\nu+\frac 
Q2\delta^\lambda{}_\nu\right)\nabla_\lambda(f_Q-f_C)
+\frac{\nabla_\nu Q}2 (f_Q-f_C)  \notag\\
&&\ \ \ \ \ 
+\left(\m R_{\mu\nu}- \m\nabla_\mu\m\nabla_\nu+\m\nabla_\nu\m\nabla_\mu\right)\nabla^\mu f_C
+\frac{\nabla_\nu\m R}2f_C\,.
\end{eqnarray}
Now, one can easily verify that 
$
-\nabla_\nu f+(\nabla_\nu Q) (f_Q-f_C)+(\nabla_\nu\m R)f_C
=0$, while   the   contracted second Bianchi identity leads 
to $
    \left(\m R_{\mu\nu}- 
\m\nabla_\mu\m\nabla_\nu+\m\nabla_\nu\m\nabla_\mu\right)\nabla^\mu f_C=0$, and 
therefore equation (\ref{vv0}) reduces to 
\begin{align}\label{eqn:divT}
\kappa \m\nabla_\mu T^\mu{}_\nu
=&\Big(2\m\nabla_\mu P^{\lambda\mu}{}_\nu
    +2P^{\lambda\mu}{}_\nu\m\nabla_\mu \Big)\nabla_\lambda(f_Q-f_C)
  +\Big(\m G^\lambda{}_\nu +\frac 
Q2\delta^\lambda{}_\nu\Big)\nabla_\lambda(f_Q-f_C)\,.
\end{align}
Furthermore, as we show in   Appendix \ref{app:divT=FE2}, the following 
identity is crucial in determining the covariant divergence of the stress-energy 
tensor:
\begin{align}\label{eqn:divT=FE2}
2(\nabla_\lambda-\tilde L_\lambda)(\nabla_\mu-\tilde L_\mu)
\left[(f_Q-f_C)P^{\lambda\mu}{}_\nu\right]
=\kappa\m\nabla_\mu T^\mu{}_\nu\,,
\end{align}
whose left-hand-side matches the expression of the connection's field equation 
(\ref{eqn:FE2-invar}) in the absence of the hypermomentum tensor.
Hence in view of (\ref{eqn:divT}), we can conclude that in $f(Q,C)$ theory the conservation of the stress energy tensor is equivalent to the affine connection's field equation, as long as the matter Lagrangian is independent of the affine connection.
Lastly, note that  in the content of the Palatini formulation, that is when the 
hypermomentum tensor is non-zero, we extract the following additional 
constraint on the $f(Q,C)$ theories
{\small{
\begin{equation}\label{eqn:divT=0}
\left(2\m\nabla_\mu P^{\lambda\mu}{}_\nu
    +2P^{\lambda\mu}{}_\nu\m\nabla_\mu
    +\m G^\lambda{}_\nu
   +\frac Q2\delta^\lambda{}_\nu\right)\nabla_\lambda(f_Q-f_C)=0\,,
\end{equation}}}
on the basis of the assumption that $Q$ and $C$ are not both constants,  and $f$ 
is not a linear function,   ensuring $(f_Q-f_C)_{,\lambda}\neq 0$.

\section{ $f(Q,C)$  cosmology}
\label{sec4}

In this section we apply $f(Q,C)$ gravity at a cosmological framework, namely 
we present $f(Q,C)$ cosmology. We consider 
 a homogenous and isotropic 
flat Friedmann-Robertson-Walker (FRW) spacetime given by the line
element in Cartesian coordinates 
\begin{equation}
ds^{2}=-dt^{2}+a^{2}(t)[dx^{2}+dy^{2}+dz^{2}],  \label{3a}
\end{equation}%
where $a(t)$ is  the scale factor, and its first time 
derivative provides the Hubble parameter 
$H(t)=\frac{\dot{a}}{a(t)}$.  

As we showed in the previous section, in the 
  the framework of $f(Q,C)$ gravity we obtain an extra, effective 
sector of geometrical origin given in (\ref{T^eff}). Thus, in a cosmological 
context, this term will correspond to an effective dark-energy sector with 
energy-momentum tensor
 \begin{equation}
   T^{\text{DE}}_{\mu\nu}= \frac 1{f_Q}\left[\frac 
f2g_{\mu\nu}-2P^\lambda{}_{\mu\nu}\nabla_\lambda(f_Q-f_C)
  -\frac {Qf_Q}2g_{\mu\nu}  -\left(\frac 
C2g_{\mu\nu}-\m\nabla_{\mu}\m\nabla_{\nu}
  +g_{\mu\nu}\m\nabla^\alpha\m\nabla_\alpha \right)f_C\right]\,.
\end{equation}

The cosmological symmetry, namely the homogeneity and isotropy,  of the FRW 
metric (\ref{3a}) can be represented by the spatial rotational and translational 
transformations.
A symmetric teleparallel affine connection is a torsion-free, curvature-free
affine connection, with both spherical and translational symmetries, implying 
that the Lie derivatives of the connection coefficients with respect to the 
generating vector fields of spatial rotations and translation vanish. There are  
three types of affine connections with such symmetries
\cite{Shi:2023kvu}. 
In order to proceed to specific cosmological applications we need to consider
  each of these symmetric teleparallel connections
and this is performed in the following subsections.

\subsection{Connection Type I
}\label{coinc}

We first consider the case with vanishing affine connection 
$\Gamma^\alpha{}_{\mu\nu}=0$   when fixing the 
coincident gauge (in general for that type of
connection this is not the case). In particular, performing a    coordinate 
transformation from the   coordinates $\tilde{x}^{\mu}$ to the 
coincident gauge   $x^{\mu}$, then     
from the condition that the connection coefficients vanish
in the coincident 
gauge we obtain that in the cosmological coordinates they become 
\cite{Hohmann:2021ast}
\begin{equation}
\tilde{\Gamma}^{\mu}{}_{\nu\rho} = \frac{\partial 
\tilde{x}^{\mu}}{\partial{x}^{\sigma}}\frac{\partial^2{x}^{\sigma}}{
\partial \tilde{x}^{\nu}\partial \tilde{x}^{\rho}}\,.
\label{eqn:connection-law}
\end{equation}
Indeed, the connection components given in \cite{Shi:2023kvu} can be easily 
recovered using (\ref{eqn:connection-law}).
We calculate 
\begin{eqnarray}
&&\m G_{\mu\nu}= -(3H^2+2\dot H)h_{\mu\nu}+3H^2u_\mu u_\nu,  \\
&& \mathring{R}= 6(2H^2+\dot{H}), 
 \\ && Q=-6H^2,
\\
&&C=\mathring{R}-Q=6(3H^2+\dot{H}),
\end{eqnarray}
where $u_\nu=(dt)_\nu$, $h_{\mu\nu}=g_{\mu\nu}+u_\mu u_\nu$. Introducing these 
into the general field equations (\ref{eqn:FE1-pre}) we   derive the 
Friedmann-like equations as
\begin{eqnarray}
&&3H^2=\kappa (\rho_m+\rho_{DE})
\label{FR1}\\
&&-(2\dot{H}+3H^2)=\kappa( p_m  +p_{DE}), 
\label{FR2}
\end{eqnarray}
where $\rho_m$ and $p_m$ are the energy density and pressure of the matter sector 
considered as a perfect fluid, and where we have defined the effective 
dark-energy density and pressure as
\begin{equation}
\rho_{DE}:=\frac{1}{\kappa}\left[ 3H^2 (1-2f_Q)
-\frac f2+(9H^2+3\dot{H})f_C-3H\dot{f_C}\right]
\label{rho}
\end{equation}
\begin{eqnarray}
&&\!\!\!\!\!\!
p_{DE}:=\frac{1}{\kappa}\Big[
  -2\dot{H}(1-f_Q)-3H^2  (1-2f_Q)
+ \frac f2  +2H\dot f_Q 
-(9H^2+3\dot{H})f_C+\ddot{f_C}\Big].
\label{p}
\end{eqnarray}
Comparing the above equations  with the modified Friedmann  equations   of 
$f(\mathbb{T},B)$ gravity under the metric teleparallelism 
\cite{teleparallel}, we observe that we have a coincidence (note also that in 
flat FRW geometry $C=B$). Thus, we conclude that 
in   this connection choice,  $f(Q,C)$ cosmology does not yield   new 
cosmological dynamics  comparing to the interesting cosmological phenomenology 
\cite{Bahamonde:2015zma,Farrugia:2018gyz,Escamilla-Rivera:2019ulu,
Paliathanasis:2017efk,Paliathanasis:2017flf,Paliathanasis:2021kuh,
Paliathanasis:2021ysb,Paliathanasis:2022pgu,
Caruana:2020szx, Moreira:2021xfe} of $f(\mathbb{T},B)$ theory  
\cite{Bahamonde:2015zma}.

Let us now proceed to the investigation of     non-vanishing 
affine connections
 in terms of Cartesian coordinates
within symmetric teleparallel class, namely with vanishing 
curvature and torsion. In 
particular, we will examine two cases. In fact, these 
two connections belong to the only two possible  classes of (non-vanishing) 
affine connections which are invariant under rotations and spatial translations 
due to the homogeneity and isotropy of
the spatially-flat FRW metric (\ref{3a}). 
 
\subsection{Connection Type II}\label{noncoinc-1}

We consider a non-vanishing affine connection 
$\Gamma$ whose non-trivial coefficients are given by  
\cite{Shi:2023kvu}
\begin{align}
   \Gamma^t{}_{tt}&=
  \gamma(t)-3H(t),    
	\quad   \Gamma^i{}_{it}=  \gamma(t)
    \quad   \Gamma^i{}_{ti}=  \gamma(t),
    \label{FormulI}
\end{align}
with ($i=1,2,3$),
which as mentioned above lead to vanishing  torsion and Riemann tensor 
components, whereas  the non-metricity tensor 
components are not all zero, giving rise to  
\begin{eqnarray}
   && Q= -6H^2+9\gamma H+3\dot\gamma,\\
   && C=\mathring{R}-Q=6(3H^2+\dot{H})-9\gamma H-3\dot\gamma.
\end{eqnarray}
In this case, substitution into the general field equations (\ref{eqn:FE1-pre}) 
gives rise to the  Friedmann   equations   
  \begin{eqnarray}
&&\!\!\!\!\!\!\!\!\!\!\!\!\!\!\!\!\!\!\!\!\!\!
\kappa \rho_m-\frac12f
=\left[6H^2-\frac32(3H\gamma+\dot\gamma) \right]f_Q
     \notag\\
     &&
 \ \ \ \ \ 
  +\left[-3\dot H-9H^2+\frac32(3H\gamma+\dot\gamma)\right]f_C \notag\\
        && 
 \ \ \ \ \  +\frac32\gamma\dot f_Q  +\frac12\left(-3\gamma+6H\right)\dot f_C,
    \end{eqnarray}
\begin{eqnarray}
&&
\!\!\!\!\!\!\!\!\!\!\!\!\!\!\!\!\!\!\!\!\!\!
\kappa p_m+\frac12f
=\left[-2\dot H-6H^2+\frac32(3H\gamma+\dot\gamma)     \right]f_Q
   \notag\\
 &&
 \ \ \ \ \ 
 +\left[3\dot H+9H^2-\frac32(3H\gamma+\dot\gamma) \right]f_C
   \notag\\  
   &&  \ \ \ \ \  
   +\frac12\left(3\gamma-4H\right)\dot f_Q  -\frac32\gamma \dot 
f_C-\ddot f_C .
    \end{eqnarray}
It proves convenient to focus on the case $9\gamma H+3\dot\gamma=0$
which 
gives   
\begin{align}
    \gamma(t)= \frac{\gamma_0}{a^{3}(t)},
    \label{specificsol1}
\end{align}
since in this case we obtain
  the same 
non-metricity scalar and boundary term as in Type I, namely
\begin{eqnarray}
\label{Qdef11}
   && Q=-6H^2,\\
   && C=\mathring{R}-Q=6(3H^2+\dot{H}).
   \label{Qdef22}
\end{eqnarray}
In this case,  the  Friedmann   equations become  (\ref{FR1}),(\ref{FR1}), but 
now the 
effective dark energy density and pressure read as
\begin{eqnarray}
&&
\!\!\!\!\!\!\!\!
\rho_{DE}:=\frac{1}{\kappa}\Big[ 
  -\frac f2-3H^2f_Q-\frac{3\gamma_0}{2a^3}\dot{f_Q}+(9H^2+3\dot{H})f_C 
-3H\dot{f_C}+\frac{3\gamma_0}{2a^3}\dot{f_C}\Big]
    \label{rho-e1}\\
&&\!\!\!\!\!\!\!\!
p_{DE}:=\frac{1}{\kappa}\Big[ \frac f2+3H^2f_Q
    +\left(2H-\frac{3\gamma_0}{2a^3}\right)\dot f_Q
        -(9H^2+3\dot{H})f_C
    +\frac{3\gamma_0}{2a^3}\dot{f_C}
    +\ddot{f_C}\Big].
\label{p-e1}
\end{eqnarray}
As we observe, this case is   different than $f(\mathbb T,B)$ cosmology. 
Finally, note 
that 
the divergence of the energy-momentum tensor $T_{\mu\nu}$ from 
(\ref{eqn:divT=FE2}) yields the modified continuity relation as
\begin{align}
 \dot{\rho}_m+3 H(\rho_m+p_m)
= \frac {3\gamma_0}{2\kappa a^3}\left[3H(\dot f_Q-\dot f_C)+(\ddot f_Q-\ddot 
f_C)\right].
\end{align}
Thus, the present scenario gives rise to an effective interaction between 
dark energy and dark matter, and such terms are known to lead to interesting 
phenomenology and alleviate the coincidence problem 
\cite{Barrow:2006hia,Amendola:2006dg,Chen:2008ft,Gavela:2009cy,Chen:2011cy,
Yang:2014gza,
Faraoni:2014vra}.

\subsection{Connection Type III }\label{nonocoinc-2}

Finally, we consider a  non-vanishing, torsion-free and curvature-free 
affine connection $\Gamma$ whose non-trivial coefficients are given by 
($i=1,2,3$) \cite{Shi:2023kvu}
\begin{align}
   \Gamma^t{}_{tt}&=-H(t),    
	\quad   \Gamma^t{}_{ii}=  \gamma(t)\,,
\end{align}
which lead to 
\begin{eqnarray}
   && Q= -6H^2+\frac{3\gamma H}{a^2}+\frac{3\dot\gamma}{a^2},\\
   && C=\mathring{R}-Q=6(3H^2+\dot{H})-\frac{3\gamma 
H}{a^2}-\frac{3\dot\gamma}{a^2}.
\end{eqnarray}
In this case, substitution into the general field equations (\ref{eqn:FE1-pre}) 
gives rise to the  Friedmann   equations   
\begin{eqnarray}
&&\!\!\!\!\!\!\!\!\!\!\!\!\!\!\!
\kappa \rho_m-\frac12f
=\left[6H^2-\frac32\frac1{a^2}(H\gamma+\dot\gamma)      \right]f_Q
      \notag\\
 & & \ \ \ \ \ \ \ \ \ 
 +\left[-3\dot H-9H^2 
+\frac32\frac1{a^2}(H\gamma+\dot\gamma) \right]f_C\notag\\
 & &
 \ \ \ \ \ \ \ \ \ 
 -\frac32\frac{\gamma}{a^2}\dot f_Q
    +\frac12\left(3\frac{\gamma}{a^2}+6H\right)\dot f_C,
\end{eqnarray}
\begin{eqnarray}
&&\!\!\!\!\!\!\!\!\!\!\!\!\!\!\!
\kappa p_m+\frac12f
=\left[-2\dot H-6H^2 +\frac32\frac1{a^2}(H\gamma+\dot\gamma)      \right]f_Q
       \notag\\
 & & \ \ \ \ \ \ \ \ \ 
 +\left[3\dot H+9H^2 -\frac32\frac1{a^2}(H\gamma+\dot\gamma) \right]f_C
   \notag\\
 & & \ \ \ \ \ \ \ \ \ 
  +\frac12\left(\frac\gamma{a^2}-4H\right)\dot f_Q 
    -\frac12\frac\gamma{a^2}\dot f_C-\ddot f_C.
\end{eqnarray}
For simplicity we focus on the case $\gamma H+\dot\gamma=0$,
which gives again    
$
    \gamma(t)= \gamma_0/a^{3}(t),
$
and thus  we acquire
  the same  non-metricity tensor  and boundary term 
(\ref{Qdef11}),(\ref{Qdef22}) as in Type I and Type II.
In this case,  the  Friedmann   equations become (\ref{FR1}),(\ref{FR1}), 
but 
now the 
effective dark energy density and pressure read as 
\begin{eqnarray}
&&\
\!\!\!\!\!\!\!\!\!\!
\rho_{DE}:=\frac{1}{\kappa}\Big[ 
   -\frac f2-3H^2f_Q+\frac{3\gamma_0}{2a^3}\dot{f_Q}+(9H^2+3\dot{H})f_C
   -\left(\frac{3\gamma_0}{2a^3}+3H\right)\Big]
   \dot{f_C}
\label{rho-e2}\\
&&\!\!\!\!\!\!\!\!
p_{DE}:=\frac{1}{\kappa}\Big[
    \frac f2+3H^2f_Q
    -\left(\frac{\gamma_0}{2a^3}-2H\right)\dot f_Q
     -(9H^2+3\dot{H})f_C
    +\frac{\gamma_0}{2a^3}\dot{f_C}
    +\ddot{f_C}\Big]
\label{p-e2}.
\end{eqnarray}
Finally, in this case 
the modified continuity equation is given by
\begin{align}
 \dot{\rho}_m+3 H(\rho_m+p_m)
=\frac{3\gamma_0}{2\kappa  a^3}\left[H(\dot f_Q-\dot f_C)-(\ddot f_Q-\ddot 
f_C) \right],
\end{align}
where as in the previous case we also obtain   an effective interaction between 
dark energy and dark matter.

\subsection{Specific example}

In this subsection we provide a numerical elaboration of a specific example. 
We use   formulation of Section \ref{noncoinc-1}, namely ansatz 
(\ref{FormulI}) 
with condition (\ref{specificsol1}). 
Hence, since we have imposed a specific form for the connection coefficients, we 
do not need to specify $f(Q,C)$ since this will arise from the solution of the 
connection equation.
As usual we focus on physically 
interesting observables such as the matter and dark energy density parameters, 
defined as
\begin{eqnarray}
&&  \Omega_m:=\frac{\kappa\rho_m}{3H^2}\\
  && \Omega_{DE}:=\frac{\kappa\rho_{DE}}{3H^2},
\end{eqnarray}
as well as on the effective dark-energy equation-of-state parameter
\begin{eqnarray}
 w_{DE}:=\frac{p_{DE}}{\rho_{DE}}.
 \label{wDE1}
\end{eqnarray}
Additionally, we will use the redshift $z$ as the independent variable.
 \begin{figure}[!]
\centering                                 
\vspace{-0.1cm}
\includegraphics[width=7.8cm]{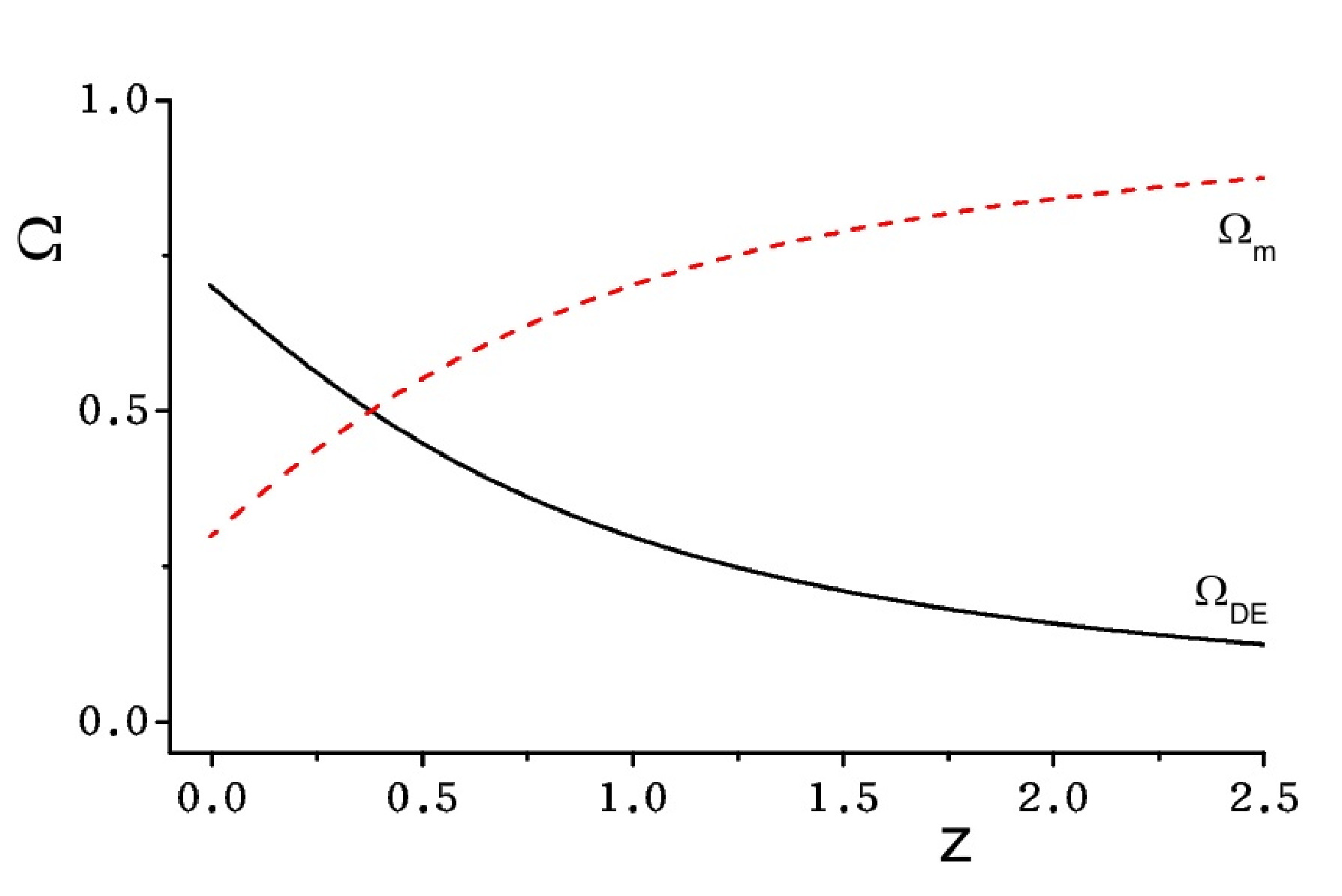} 
\caption{\it{  The evolution of the effective dark energy density 
parameter $\Omega_{DE}$   and of the matter density parameter 
$\Omega_{m}$,   as a function of the redshift $z$, for   $f(Q,C)$ cosmology 
with    Type II non-vanishing connection   (\ref{FormulI}) with condition 
(\ref{specificsol1}),   
with $\gamma_0=1$ in units 
where 
$\kappa=1$. We have imposed $ \Omega_m(z=0):= \Omega_{m0}=0.31$ in agreement 
 with observations \cite{Planck:2018vyg}.
}}
\label{fig1}
\end{figure}

\begin{figure}[htbp]
 \vspace{-0.1cm}
\centering                                 
\includegraphics[width=8.3cm]{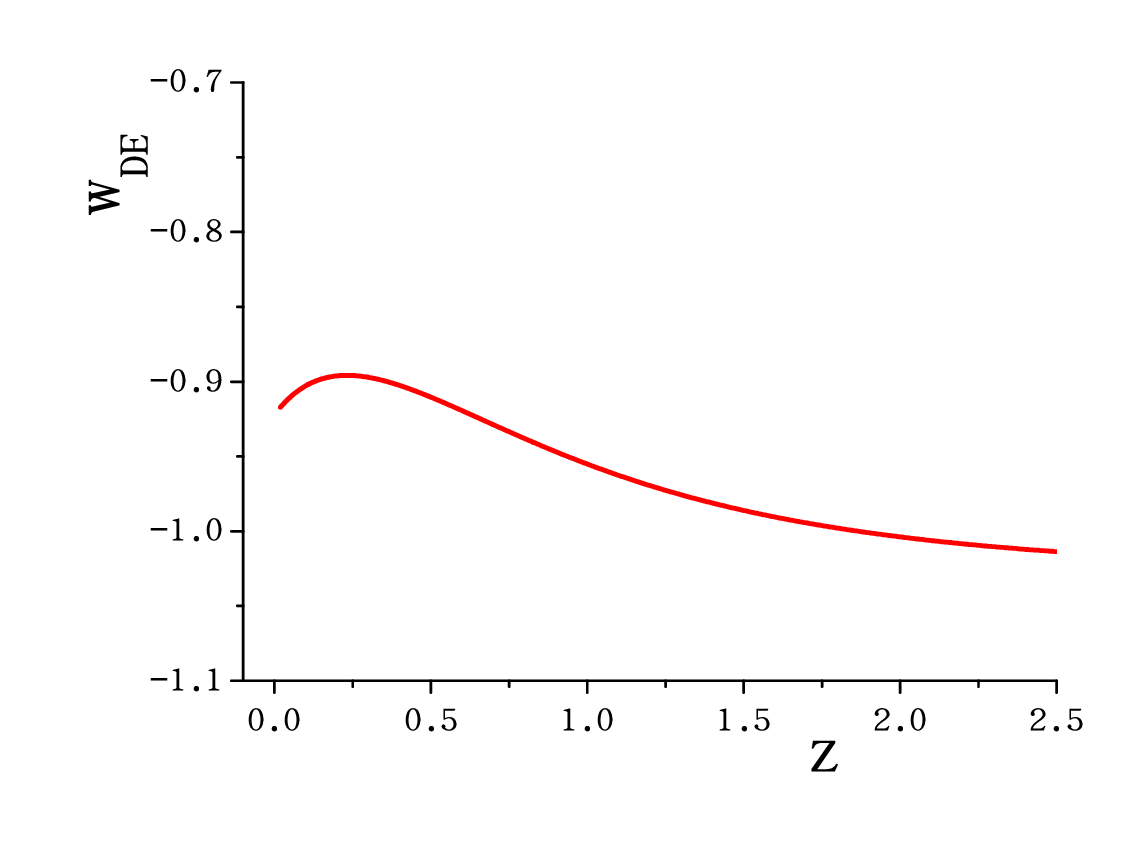} 
\caption{\it{  The evolution of the effective dark-energy equation-of-state 
parameter  $w_{DE}$ given in (\ref{wDE1}),   as a function of the redshift $z$, 
for   $f(Q,C)$ cosmology 
with  with    Type II non-vanishing connection     (\ref{FormulI}) with 
condition 
(\ref{specificsol1}),  
with $\gamma_0=1$ in units 
where 
$\kappa=1$. We have imposed $ \Omega_m(z=0):= \Omega_{m0}=0.31$ in agreement 
 with observations \cite{Planck:2018vyg}.
}}
\label{fig2}
\end{figure}

We evolve equations  the Friedmann equation (\ref{FR1}), (\ref{FR2}) 
numerically and in
Fig. \ref{fig1}
we depict the evolution of $\Omega_m(z)$ and $\Omega_{DE}(z)$. As we observe, 
we obtain the usual thermal history of the universe, with the sequence of 
matter and dark-energy epochs. Additionally, in Fig. \ref{fig2} we present 
the corresponding effective dark-energy equation-of-state 
parameter  $w_{DE}(z)$. Interestingly enough we see that the effective dark 
energy crosses slightly into the phantom regime during the  evolution, a 
feature that shows the capabilities of the scenario. Indeed according to the 
definitions of the effective dark-energy density and pressure given above, one 
can deduce that $w_{DE}$ can be quintessence-like, phantom-like, or exhibit the 
phantom-divide crossing during evolution.

\section{Conclusions}\label{sec5}

In the symmetric teleparallel formulation of gravity one uses non-metricity 
instead of curvature. Since the non-metricity scalar $Q$ differs from the 
standard Levi-Civita Ricci scalar $\m R$   by a 
total divergence term, namely  $\m R=Q+C$,  general relativity and symmetric 
teleparallel general relativity are equivalent at the level of equations. 
However, modifications of the form   $f(\m R)$ and $f(Q)$ are not equivalent, 
since they do not differ by a total derivative. 

In the present work we formulated $f(Q,C)$ gravity and cosmology, by 
incorporating the boundary term $C$ alongside $Q$ in the Lagrangian. First we 
have extracted the general field equations, and then we have applied them in a 
cosmological framework, namely to flat Friedmann-Robertson-Walker (FRW) metric. 
Making 
three connection
choices, we finally obtained the 
 corresponding  modified Friedmann equations. 
As we showed, we acquired an effective dark-energy sector of geometrical 
origin, which can lead to interesting cosmological phenomenology. Additionally, 
one may obtain an effective interaction between matter and dark energy.

Using a specific example, we showed that  we can   obtain the usual thermal 
history of the universe, with the sequence of matter and dark-energy epochs, as 
required. Furthermore, as we saw, the   effective dark-energy equation-of-state 
parameter  can be quintessence-like, phantom-like, or cross the phantom-divide, 
features that show the capabilities of the scenario.

  In summary, we saw that $f(Q,C)$ gravity is a novel  class of 
gravitational modification, and its cosmological application leads to 
interesting features. However, there are many investigations that one should do 
before considering   $f(Q,C)$ gravity as a candidate for the description of 
Nature. One should proceed to a detailed confrontation with data from  
Supernovae type Ia (SNIa), Baryonic Acoustic Oscillations (BAO), Cosmic 
Microwave Background Radiation (CMB),  Cosmic Chronometer (CC) etc 
observations, in order to extract constrains on the involved functions and 
parameters, and examine whether the model can alleviate the $H_0$ tension. 
Moreover, one could try to perform a dynamical-system analysis, in order to 
bypass the complexities of the scenario and extract the global features of the 
evolution. Additionally, one could proceed to the investigation of perturbations 
in order to examine the $\sigma_8$ tension as well as the effect on 
gravitational waves,  
however  it should be pointed out that the theoretical framework for the 
description of   perturbations in a general non-Riemannian geometry requires 
to extend the standard cosmological perturbation theory in the lines of  
\cite{Aoki:2023sum}, which in the case of symmetric 
teleparallel 
gravities may lead to perturbation spectrum reduction 
\cite{Heisenberg:2023wgk}.  
 These necessary studies lie beyond the scope of the present work, 
and are left for future projects.

\subsection*{Acknowledgements}
The research was supported by the Ministry of Higher Education 
(MoHE), through the Fundamental Research Grant Scheme 
(FRGS/1/2021/STG06/UTAR/02/1).
ENS acknowledges the 
contribution of the LISA CosWG, and of   COST Actions  CA18108 
``Quantum Gravity 
Phenomenology in the multi-messenger approach''  and 
CA21136 ``Addressing observational tensions in cosmology with systematics and 
fundamental physics (CosmoVerse)''.

\appendix

\section{ Derivation of field equations }\label{derive}

In this appendix we provide the derivation of the field equations 
(\ref{eqn:FE1-pre})--(\ref{eqn:FE2-invar}).
The coincident gauge shall be taken while deriving (\ref{eqn:FE1-pre}), 
so that partial derivatives and covariant derivatives can be used 
interchangeably. As usual, throughout the appendix all the divergence terms 
contributing to the boundary terms of the integrals will be neglected during the 
derivations.


In order to derive (\ref{eqn:FE1-pre}), we perform the variation on the action 
(\ref{eqn:action-fQC}) with respect to the metric to  obtain  
{\small{\begin{eqnarray}\label{eqn:B1}
&&
\!\!\!\!\!\!\!\!\!\!\!\!\!\!
0=\delta_g S\nonumber\\
&&\!\!\!\!\!\!\!\!\!\!\!\!\!\!
=\frac1{2\kappa}\!\int \!\!\sqrt{-g}\left[\! -\frac 
f2g_{\mu\nu}\delta g^{\mu\nu}
\!    +\!f_Q\delta_gQ\!+\!f_C\delta_gC\!-\!\kappa T_{\mu\nu}\!\right] \!\!
\,d^{4}x.
\end{eqnarray}}}
The variation of $Q$ is given in the the following identity \cite{fQT}
\begin{align}
\delta_gQ=(-2P^\lambda{}_{\mu\nu}\nabla_\lambda+P_{\mu\alpha\beta}Q_\nu{}^{\alpha\beta}
-2P_{\alpha\beta\nu}Q^{\alpha\beta}{}_\mu)\delta g^{\mu\nu}\,,
\end{align}
while  the second term on the right-hand side of (\ref{eqn:B1}),
after neglecting the divergence term, reads 
\begin{eqnarray}
&&
\!\!\!\!\!\!\!\!\!\!\!\!\!\!\!\!\!\!\!\!\!
\sqrt{-g}f_Q\delta_g Q
=\sqrt{-g}\left[2\frac{\partial_\lambda(\sqrt{-g}f_QP^{\lambda}{}_{\mu\nu})}{
\sqrt{-g}}  
+
\left(P_{\mu\alpha\beta}Q_\nu{}^{\alpha\beta}
-2P_{\alpha\beta\nu}Q^{\alpha\beta}{}_\mu\right)f_Q\right]\delta g^{\mu\nu}
\,.
\label{eqn:2nd}
\end{eqnarray}
On the other hand
\begin{eqnarray}\label{fC-dC}
&&
\!\!\!\!\!\!
\!\!\!\!\!\! \sqrt{-g}f_C(\delta_gC)
=\left[-f_CC+(\nabla_\alpha f_C)(Q^\alpha-\tilde Q^\alpha)\right]
   \delta_g\sqrt{-g} 
   +\sqrt{-g}(\nabla_\alpha f_C)\delta_g(Q^\alpha-\tilde Q^\alpha)\,.
\end{eqnarray}
Since 
$
\delta_g Q_{\lambda\mu\nu}=
(-Q_{\lambda\alpha\mu}g_{\beta\nu}-g_{\alpha\mu}Q_{\lambda\beta\nu}-g_{\alpha\mu}g_{\beta\nu}\nabla_\lambda)\delta g^{\alpha\beta}\,,
$
we have 
\begin{align}
 \delta_g(Q^\alpha-\tilde Q^\alpha)
 =&\delta_g\left[(Q_{\lambda\mu\nu}
   -Q_{\mu\nu\lambda})g^{\lambda\alpha}g^{\mu\nu}\right] \nonumber\\
 =&\left[-Q^\alpha{}_{\mu\nu}
    -g_{\mu\nu}\nabla^\alpha+\delta^\alpha{}_\mu(Q_\nu+\nabla_\nu)
 \right]\delta g^{\mu\nu}\,.
\end{align}
Substituting this into (\ref{fC-dC}), gives
\begin{eqnarray}
&&\!\!\!\!\!\!\!\!\!\!\!\!
\sqrt{-g}f_C(\delta_gC) 
=\sqrt{-g}\Big[\frac C2g_{\mu\nu}f_C
   -\frac{Q^\alpha-\tilde Q^\alpha}2g_{\mu\nu}\nabla_\alpha f_C
   -Q^\alpha{}_{\mu\nu}\nabla_\alpha f_C+Q_\nu\nabla_\mu f_C
\Big]\delta g^{\mu\nu} \notag\\
 &&  \ \ \ \ \ \ \ \ \ \ \  \ \ \     
+\sqrt{-g}\left[-g_{\mu\nu}(\m\nabla^\alpha 
f_C)\nabla_\alpha
        +(\m\nabla_\mu f_C)\nabla_\nu
\right]\delta g^{\mu\nu}\,.
\end{eqnarray}
Moreover, we derive 
{\small{
\begin{eqnarray}
&&\!\!\!\!\!\!\!\!\!\!
-\sqrt{-g}g_{\mu\nu}(\m\nabla^\alpha f_C)\nabla_\alpha\delta g^{\mu\nu} \notag\\
&&\!\!\!\!
=\sqrt{-g}(\nabla_\alpha-\tilde L_\alpha)
        (g_{\mu\nu}\m\nabla^\alpha f_C)\delta g^{\mu\nu} \notag\\
&&\!\!\!\!=\sqrt{-g}(Q_{\alpha\mu\nu}\m\nabla^\alpha 
f_C+g_{\mu\nu}\m\nabla_\alpha\m\nabla^\alpha f_C)\delta 
g^{\mu\nu}\sqrt{-g}(\m\nabla_\mu f_C)\nabla_\nu\delta g^{\mu\nu}\nonumber \\
&&\!\!\!\!
 =\sqrt{-g}(-\m\nabla_\nu\m\nabla_\mu f_C+L^\alpha{}_{\mu\nu}\m\nabla_\alpha 
f_C
+\tilde L_\nu\m\nabla_\mu f_C)\delta g^{\mu\nu}\,.
\end{eqnarray}}}
By using the fact that $Q_\nu=-2\tilde L_\nu$, and (\ref{P}),
we obtain
\begin{eqnarray}
&&
\!\!\!\!\!\!\!\!\!\!\!\!\!\!\!\!\!\!\! 
\sqrt{-g}f_C\delta_gC
=\sqrt{-g}\Big[\frac C2g_{\mu\nu}f_C+g_{\mu\nu}\m\nabla_\alpha\m\nabla^\alpha 
f_C  
     -\m\nabla_\nu\m\nabla_\mu f_C
  -2P^\alpha{}_{\mu\nu}\m\nabla_\alpha f_C\Big]
  \delta g^{\mu\nu}\,.
\label{eqn:3rd}
\end{eqnarray}
Finally, Eq. (\ref{eqn:FE1-pre}) can   be obtained after substituting 
(\ref{eqn:2nd}) and (\ref{eqn:3rd}) into (\ref{eqn:B1}).

Now, we apply the method of Lagrange multipliers to derive the connection field 
equation (\ref{eqn:FE2-invar}). In particular, we write
\begin{align}
S_1=\frac1{2\kappa}\int \sqrt{-g}\left[
r_\lambda{}^{\mu\alpha\nu}R^\lambda{}_{\mu\alpha\nu}+
t_\lambda{}^{\mu\nu}\mathbb T^\lambda{}_{\mu\nu}
\right] %
\,d^{4}x\,,
\end{align}
which implements both the torsion-free and curvature-free constraints
is added to the action  (\ref{eqn:action-fQC}).
Variation with respect to the affine connection reads 
\begin{equation}
0=\delta_\Gamma S=
\frac1{2\kappa}\int \sqrt{-g}\Big[(f_Q-f_C)\delta_\Gamma Q
-\kappa\Delta_\lambda{}^{\mu\nu}\delta\Gamma^\lambda{}_{\mu\nu}
+r_\lambda{}^{\mu\alpha\nu}\delta_\Gamma R^\lambda{}_{\mu\alpha\nu}
+t_\lambda{}^{\mu\nu}\delta_\Gamma\mathbb T^\lambda{}_{\mu\nu}
\Big] %
\,d^{4}x\,,
\label{eqn:action-connection}\end{equation}
where we use  the fact that the Ricci scalar $\m R$ depends only on the metric 
components.
We calculate each term separately as follows:
\begin{align}
\delta_\Gamma Q=&-4P^{\nu\mu}{}_\lambda\delta \Gamma^\lambda{}_{\mu\nu}\,,
\end{align}
\begin{eqnarray}
&& 
\!\!\!\!\!\!\!\!\!
\sqrt{-g}r_\lambda{}^{\mu\alpha\nu}\delta_\Gamma R^\lambda{}_{\mu\alpha\nu}
 =\sqrt{-g}r_\lambda{}^{\mu\alpha\nu}
    (\nabla_\alpha\delta\Gamma^\lambda{}_{\mu\nu}-
     \nabla_\nu\delta\Gamma^\lambda{}_{\mu\alpha})\nonumber\\
     && \ \ \ \ \ \ \ \  \ \   \ \ \ \  \ \   \ \ \ \ \
 =2\sqrt{-g}(\nabla_\alpha-\tilde L_\alpha)r_\lambda{}^{\mu[\nu\alpha]}\delta 
\Gamma^\lambda{}_{\mu\nu},
\end{eqnarray}
\begin{align}
 \sqrt{-g}t_\lambda{}^{\mu\nu}\delta_\Gamma\mathbb T^\lambda{}_{\mu\nu}
 =&-2\sqrt{-g}t_\lambda{}^{[\mu\nu]}\delta \Gamma^\lambda{}_{\mu\nu}\,.
\end{align}
After substituting these equations into (\ref{eqn:action-connection}) one arrives at 
\begin{align}
Y^{\nu\mu}{}_\lambda=2(\nabla_\alpha-\tilde L_\alpha)r_\lambda{}^{\mu[\nu\alpha]}
-2t_\lambda{}^{[\mu\nu]}\,,
\end{align}
where we denote $Y^{\nu\mu}{}_\lambda=4(f_Q-f_C)P^{\nu\mu}{}_\lambda+\kappa
\Delta_\lambda{}^{\mu\nu}$.
The Lagrange multiplier terms $t_\lambda{}^{\mu\nu}$  can be eliminated if the 
symmetric part of the equation in the indices $\mu$ and $\nu$ is considered, 
which reads
\begin{align}
Y^{(\nu\mu)}{}_\lambda=(\nabla_\alpha-\tilde L_\alpha)(r_\lambda{}^{\mu[\nu\alpha]}+r_\lambda{}^{\nu[\mu\alpha]})\,.
\end{align}
Finally if we apply the operator $(\nabla_\nu-\tilde L_\nu)(\nabla_\mu-\tilde 
L_\mu)$ to this equation and then follow the calculations as in 
\cite{Hohmann:2021fpr}, 
the Lagrange multiplier  terms $r_\lambda{}^{\nu\mu\alpha}$
will be  removed and we result to
\begin{align}
(\nabla_\nu-\tilde L_\nu)(\nabla_\mu-\tilde L_\mu)Y^{(\nu\mu)}{}_\lambda=0\,.
\end{align}
Finally, taking into account of the symmetry of $(\nabla_\nu-\tilde 
L_\nu)(\nabla_\mu-\tilde L_\mu)$, we extract the connection field equation 
(\ref{eqn:FE2-invar}).

\section{ Proof of relation (\ref{eqn:divT=FE2})}\label{app:divT=FE2}

In this appendix we extract the useful relation (\ref{eqn:divT=FE2}).
Indeed it can arise as an immediate consequence of the following identity
\begin{eqnarray}
&& \!\!\!\!\!\!\!\!\!\!\!\!\!\!\!\!
 \left(2\m\nabla_\mu P^{\lambda\mu}{}_\nu+\m G^\lambda{}_\nu+\frac 
Q2\delta^\lambda{}_\nu\right) \nabla_\lambda\zeta 
 +2P^{\lambda\mu}{}_\nu\m\nabla_\mu\nabla_\lambda\zeta
  =2(\nabla_\lambda-\tilde L_\lambda)(\nabla_\mu-\tilde L_\mu)
 (\zeta P^{\lambda\mu}{}_\nu)\,,
 \label{eqn:C1}
\end{eqnarray}
where $\zeta$ is an arbitrary scalar field.
Firstly, the left-hand-side of (\ref{eqn:C1}) can be expanded as 
\begin{eqnarray}
&&
\!\!\!\!\!\!\!\!\!\!\!\!\!\!\!\!\!\!\!\!\!\!\! 
2(\nabla_\lambda-\tilde L_\lambda)(\nabla_\mu-\tilde L_\mu)
 (\zeta P^{\lambda\mu}{}_\nu)=  2\zeta(\nabla_\lambda-\tilde 
L_\lambda)(\nabla_\mu-\tilde L_\mu)
 P^{\lambda\mu}{}_\nu\nonumber\\
 &&\ \ \ \ \ \ \ \ \ \ \ \ \ \ \ \ \ \ \ \ \ \ \ \ \ \ \ \ \ \ \ 
 +2[(\nabla_\mu-\tilde L_\mu)(P^{\lambda\mu}{}_\nu+P^{\mu\lambda}{}_\nu)]
 \nabla_\lambda\zeta  
 +2P^{\lambda\mu}{}_\nu\nabla_\mu\nabla_\lambda\zeta\,.
\end{eqnarray}
As shown in \cite{avik/cqg} one has  
\begin{align}
 (\nabla_\lambda-\tilde L_\lambda)(\nabla_\mu-\tilde L_\mu)
 P^{\lambda\mu}{}_\nu=0,
\end{align}
which implies that the first term on the right-hand-side 
vanishes.
Additionally, we use a second identity  given in  \cite{avik/cqg}, namely
\begin{align}
2(\nabla_\lambda-\tilde L_\lambda)P^{\lambda\alpha}{}_\nu
=\m G^\alpha{}_\nu+\frac Q2\delta^\alpha{}_\nu
    +2L^\sigma{}_{\nu\beta}P^{\alpha\beta}{}_\sigma\,,
\end{align}
while  a direct calculation gives
\begin{equation}
 2(\nabla_\mu-\tilde L_\mu)P^{\lambda\mu}{}_\nu
=2\m\nabla_\mu P^{\lambda\mu}{}_\nu+2L^\lambda{}_{\sigma\mu}P^{\sigma\mu}{}_\nu
 -2L^\sigma{}_{\nu\mu}P^{\lambda\mu}{}_\sigma\,,
\end{equation}
 and 
 \begin{equation}
2P^{\lambda\mu}{}_\nu\nabla_\mu\nabla_\lambda\zeta
=2P^{\lambda\mu}{}_\nu\m\nabla_\mu\nabla_\lambda\zeta
    -2P^{\lambda\mu}{}_\nu L^\sigma{}_{\lambda\mu}\nabla_\sigma\zeta\,.
\end{equation}
Summing these equations, one   arrives at (\ref{eqn:C1}).
Hence, by virtue of (\ref{eqn:divT}) and (\ref{eqn:C1}), we finally obtain 
relation (\ref{eqn:divT=FE2}).
 

\end{document}